\documentclass[aps,prb,groupedaddress,twocolumn,showpacs]{revtex4-1}
\usepackage{amsmath}
\usepackage{amssymb}
\usepackage{graphicx}
\usepackage[usenames,dvipsnames]{xcolor}
\usepackage{tikz}
\usepackage{pgffor}
\usepackage{verbatim}
\usepackage{bbm}
\usepackage{float}
\usepackage{mathrsfs}
\usepackage[colorlinks, breaklinks=true,linkcolor=red, citecolor=blue, linktocpage=true]{hyperref}

\newcommand{\modulo}[1]{\ ({\rm mod} \ #1)}

\begin{document}

\title{Braiding statistics and classification of two-dimensional charge-$2m$ superconductors}
\author{Chenjie Wang}
\affiliation{Perimeter Institute for Theoretical Physics, Waterloo, Ontario N2L 2Y5, Canada}
\date{\today}
\begin{abstract}
We study braiding statistics between quasiparticles and vortices in two-dimensional charge-$2m$ (in units of $e$) superconductors that are coupled to a $\mathbb Z_{2m}$ dynamical gauge field, where $m$ is any positive integer. We show that there exist $16m$ types of braiding statistics when $m$ is odd, but only $4m$ types when $m$ is even.  Based on the braiding statistics, we obtain a classification of topological phases of charge-$2m$ superconductors---or formally speaking, a classification of symmetry-protected topological phases, as well as invertible topological phases, of two-dimensional gapped fermions with $\mathbb Z_{2m}^f$ symmetry. Interestingly, we find that there is no nontrivial fermionic symmetry-protected topological phase with $\mathbb Z_4^f$ symmetry.
\end{abstract}



\maketitle

\section{Introduction}

Since the discovery of topological insulators, much attention has been attracted to the interplay between symmetry and topology in gapped quantum-many body systems.\cite{hasan10,qi11} A natural and rich generalization of topological insulators is the so-called symmetry-protected topological (SPT) phases, which are systems with an energy gap and a (general) symmetry that is not spontaneously broken. Importantly, SPT systems do not support exotic bulk excitations (with fractional statistics or fractional quantum number), nevertheless they support nontrivial edge/surface states that are protected by the symmetry.\cite{chen13,senthil15}

Recently a great advance was achieved in the classification and characterization of bosonic SPT phases. \cite{gu09, pollmann10, fidkowski11,chen11a, chen11b, schuch11,chen13,senthil15} Bosonic SPT phases intrinsically require strong interaction to support an energy gap.   On the other hand, SPT phases of interacting fermions are less understood. Most of the understanding results from the study of interaction effects on the free-fermion classification\cite{ff1,ff2,fidkowski10, gu14b, ryu12,yao13,qi13,wangc-science, fidkowski13,metlitski14,wangc14,you14,morimoto15,queiroz16,witten15}. Fewer works have been done starting with interacting fermions.\cite{gu-super,kapustin14,cheng15} Moreover, most of the latter works focus on symmetries of the form $G_f= \mathbb Z_2^f \times G$, where $\mathbb Z_2^f$ is the group of fermion parity. That is, $G_f$ is a trivial $\mathbb Z_{2}^f$ group extension of $G$. Nevertheless, $G_f$ can generally be any $\mathbb Z_2^f$ extension of $G$. It remains an open question how to classify fermionic SPT phases with general $G_f$.

In this work, we study the classification of two-dimensional interacting fermionic SPT phases with $G_f=\mathbb Z_{2m}^f$ symmetry, which is the simplest example with $G_f$ being a nontrivial $\mathbb Z_2^f$ extension.\footnote{When $m$ is odd, $\mathbb Z_{2m}^f$ remains a trivial $\mathbb Z_2^f$ extension because of the isomorphism $\mathbb Z_{2m}^f = \mathbb Z_2^f\times\mathbb Z_m$.} Physically, such fermionic systems can be thought of as {\it charge-$2m$ superconductors}\cite{berg09, radzihovsky09,herland10,agterberg11,moon12}, where a cluster of $2m$ fermions condense, making the fermion number be conserved only modulo $2m$. We achieve a classification of charge-$2m$ superconductors in two steps. First, we argue that by coupling charge-$2m$ superconductors to a $\mathbb Z_{2m}$ dynamical gauge field, braiding statistics in the resulting {\it gauged superconductors} come only in $16m$ types when $m$ is odd, and $4m$ types when $m$ is even. The method of gauging symmetry has been applied previously to study various bosonic SPT phases\cite{levin12,wangcj15,threeloop} as well as fermionic SPT phases.\cite{gu14b} Second, from the classification of braiding statistics, we deduce a classification of topological phases of charge-$2m$ superconductors (Table \ref{tab1}). We hope that the current study of $\mathbb Z_{2m}^f$ fermionic systems can inspire future studies on general fermionic SPT phases.



The rest of the paper is organized as follows. In Sec.~\ref{sec2}, we derive a classification of braiding statistics in gauged charge-$2m$ superconductors. To support the classification, we construct models to realize all types of braiding statistics in Sec.~\ref{sec3}. Based on the braiding statistics, a classification of 2D fermionic SPT phases, as well as invertible topological phases, with $\mathbb Z_{2m}^f$ symmetry, is deduced in Sec.~\ref{sec4}. We conclude in Sec.~\ref{conclusion}. The appendices contain several technical details.


\section{Braiding statistics in gauged charge-$2m$ superconductors}
\label{sec2}

In this section, we derive all types of braiding statistics in gauged charge-$2m$ superconductors.  The case that $m=1$ (i.e., regular charge-2 superconductors) was studied before by Kitaev, and it was found that there are 16 types of distinct braiding statistics.\cite{kitaev06} That result was derived by directly solving the pentagon and hexagon equations, which are equations that any braiding statistics should satisfy. In this work, we use a more physical way to derive the braiding statistics in $\mathbb Z_{2m}$ discrete gauge theories coupled to fermionic matter.

\subsection{Basic properties}

We begin by discussing excitations in $\mathbb Z_{2m}$ discrete gauge theories coupled to charge-$2m$ superconductors. We will call such systems {\it gauged superconductors}. We consider a discrete gauge field in a gapped and deconfined phase.\footnote{This can be guaranteed by choosing a proper gauging procedure, see Refs.~\cite{levin12,wangcj15}.} Excitations in gauged superconductors can be divided into two types: \emph{charges} (i.e., Bogoliubov quasiparticles) and \emph{vortices}. There are $2m$ charge excitations in total, which carry gauge charges $0, 1, \dots, (2m-1)$ respectively. (We set $e=\hbar=c=1$ throughout this paper.) An important consequence of the fact that charge-$2m$ superconductors are made out of fermions is that: if a charge $q$ carries an odd number of gauge charge, it is a fermion; otherwise, it is a boson. Therefore, the exchange statistics $\theta_q$ is given by
\begin{equation}
\theta_q = \pi q \label{exchange}
\end{equation}
where we have used $q$ to denote both the excitation and its gauge charge.

Vortices are excitations that carry nonvanishing gauge flux $\frac{\pi}{m}k$, where $k=1, \dots, (2m-1)$. Unlike charges, vortices are not uniquely labeled by their gauge flux. A vortex $\alpha$ may have the same flux as another vortex $\alpha'$, but differ from $\alpha'$ by attaching some amount of charge. In fact, every other vortex carrying the same flux as $\alpha$ can be obtained from $\alpha$ by attaching some charge.\cite{wangcj15}
The statistical phase $\theta_{q\alpha}$ associated with braiding a charge $q$ around a vortex $\alpha$ should follow the Aharonov-Bohm law:
\begin{equation}
\theta_{q\alpha} = q \phi_\alpha \label{ab_phase}
\end{equation}
where $\phi_\alpha$ is the gauge flux carried by $\alpha$.

The above properties have almost defined our system: Eq.~(\ref{exchange}) shows that the matter is fermionic, and Eq.~(\ref{ab_phase}) shows that the gauge group is $\mathbb Z_{2m}$. One can see that the only undetermined braiding statistics are those between two vortices, which in general can be non-Abelian.


\subsection{Fusion rules}

To find possible braiding statistics between vortices in gauged charge-$2m$ superconductors, we first study fusion rules between vortices. We are not interested in general fusion rules, but only a particular one: the fusion rule between $\xi$ and its anti-particle $\bar\xi$, where $\xi$ is an arbitrary vortex carrying {\it unit flux} $\frac{\pi}{m}$. This fusion rule turns out to capture some key features of gauged charge-$2m$ superconductors.

The fusion rule between $\xi$ and $\bar\xi$ can be generally written as
\begin{equation}
\xi \times \bar\xi= 0+  q + \dots, \label{fusion1}
\end{equation}
where $0$ is the  vacuum  anyon, $q$ is some charge, and ``$\dots$'' represents any other charges. Only charges appear on the right-hand side of (\ref{fusion1}), because $\xi$ and $\bar\xi$ carry opposite gauge flux and accordingly any fusion channel should carry zero flux.  We have taken the fusion multiplicity $N_{\xi\bar\xi}^q$ to be $1$, which can be proven but is not essential for the following discussion (see Ref.~\onlinecite{wangcj15} for a derivation).

We now use Eqs.~(\ref{exchange}) and (\ref{ab_phase}) to constrain what charges are allowed on the right-hand side of  (\ref{fusion1}).  First, we notice the following formula from general algebraic theory of anyons \cite{kitaev06}:
\begin{align}
R_{\bar\xi\xi}^q R_{\xi\bar\xi}^q & = e^{i(\theta_q-\theta_\xi-\theta_{\bar\xi})} \label{topospin}
\end{align}
where $R_{\xi\bar\xi}^q$ is the $R$ symbol associated with a half-braiding between $\xi$ and $\bar\xi$ in the fusion channel $q$, and $\theta_q,\theta_\xi,\theta_{\bar\xi}$ are topological spins\footnote{We use $\theta_\alpha$, instead of $s_\alpha \equiv \theta_{\alpha}/2\pi$, as the topological spin to avoid repetitive appearance of $2\pi$ in many formulas.}. (For Abelian anyons, topological spin is the same as exchange statistics.) In addition, we show in Appendix \ref{app1} that the mutual statistics $R_{\bar\xi\xi}^q R_{\xi\bar\xi}^q$ between $\xi$ and $\bar \xi$ in the channel $q$ also satisfies the following relation
\begin{equation}
 R_{\bar\xi\xi}^q R_{\xi\bar\xi}^q = R_{\bar\xi\xi}^0 R_{\xi\bar\xi}^0\ e^{i\frac{\pi q}{m}}  \label{braid1}
\end{equation}
Combining (\ref{topospin}) and (\ref{braid1}), we obtain the following constraint on $q$:
\begin{equation}
e^{i \frac{\pi q}{m}} = e^{i\theta_q}
\label{constraint}
\end{equation}

Using Eq.~(\ref{exchange}), we find that the solutions to (\ref{constraint}) depend on the parity of $m$: If $m$ is even, $q$ can only be $0$; if $m$ is odd, $q$ can be $0$ or $m$.  Therefore,  possible fusion rules for $\xi$ and $\bar\xi$ are:
\begin{align}
\text{$m$ being even:} & \quad \quad\quad\quad \xi\times\bar\xi = 0 \label{fusion3}\\
\text{$m$ being odd:} \ & \quad \xi\times\bar\xi = 0\ \ \text {or} \ \ \xi\times\bar\xi = 0 + Q \label{fusion2}
\end{align}
where we have used $Q$ to denote the special charge that carries gauge charge $m$. Note that $Q$ is a fermion when $m$ is odd.


\subsection{ $m$ being even }

With the above fusion rules, we now study braiding statistics between vortices. We start with the simpler case of even $m$. We show that there are $4m$ types of braiding statistics in this case.

With the fusion rule (\ref{fusion3}) for even $m$, we first claim that all excitations in the system are Abelian anyons. To see that, we notice that since $\xi$ carries unit flux, a general vortex can be obtained by fusing a number of $\xi$'s and some charge. We also observe that: (1) $\xi$ is Abelian because of (\ref{fusion3}); (2) any charge is Abelian; (3) fusing two Abelian anyons produces a new Abelian anyon. Combining all together, we prove the claim.

Next, we define the following quantity
\begin{equation}
\Theta = 2m\theta_\xi \label{invariant}
\end{equation}
where $\theta_\xi$ is the exchange statistics of $\xi$. We will name $\Theta$ as {\it topological invariant}, following the terminology of Ref.~\onlinecite{wangcj15}, where similar quantities were defined. The topological invariant $\Theta$ holds two nice properties: (i) it only depends on the flux of $\xi$; (ii) the full set of braiding statistics can be reconstructed out of $\Theta$. The second point is particularly useful because the information contained in braiding statistics is now summarized in a single quantity $\Theta$.

To see point (i), we replace $\xi$ with $\xi'$ in the definition (\ref{invariant}), where $\xi'$ also carries unit flux. Using the fact that $\xi, \xi'$ only differ by some charge, one can show that the difference in $\Theta$ is always a multiple of $2\pi$. Hence, only the flux of $\xi$ matters for $\Theta$.  To show point (ii), let us determine the full set of exchange and mutual statistics from a given $\Theta$.  The definition (\ref{invariant}) implies that $\theta_\xi$ can be generally written as $\theta_\xi = \frac{\Theta}{2m} + \frac{\pi }{m}(\text{integer})$. By attaching charge to $\xi$, we find that there always exists a vortex $\xi_0$ such that its exchange statistics $\theta_{\xi_0} = \frac{\Theta}{2m}$. Then, a general excitation can be obtained by fusing $k$ copies of $\xi_0$'s and further fusing a charge $q$, which we label as $(\xi_0; k, q)$. One may refer to $\xi_0$ as the {\it reference vortex}. Using (\ref{exchange}), (\ref{ab_phase}), $\theta_{\xi_0} = \frac{\Theta}{2m}$ and properties of Abelian anyons, it is not hard to show that the exchange statistics of $(\xi_0; k, q)$ is given by
\begin{equation}
\theta_{(\xi_0;k,q)} = \frac{\Theta}{2m}k^2  + \frac{\pi }{m} k q + \pi q \label{spin1}
\end{equation}
Mutual statistics can be similarly determined but we do not list them here. One can see that there are $4m^2$ distinct excitations in total, labeled by $(\xi_0; k, q)$ with $k, q$ in the range $0,\dots,(2m-1)$.

Having established the two properties of $\Theta$, all that remains is to find possible values of $\Theta$. To do that, we imagine braiding a vortex $\xi$ around a collection of $2m$ other $\xi$'s. The statistical phase associated with this braiding process  can be easily computed, and is given by $2m\cdot 2\theta_\xi =2\Theta$. On the other hand, the collection of $2m$ $\xi$'s can be fused into a charge, implying that the total statistical phase also equals a  multiple of $\frac{\pi}{m}$. Therefore, we are led to the following constraint on $\Theta$:
\begin{equation}
\Theta = \frac{\pi}{2m} p \label{constraint1}
\end{equation}
where $p$ is an integer. For $p=0, 1, \dots, 4m-1$, we find $4m$ distinct values of $\Theta$. We will argue in Sec.~\ref{sec3} that all these values of $\Theta$ can be realized in physical systems. Therefore, we find $4m$ distinct types of braiding statistics in total, all of which are Abelian.


\subsection{$m$ being odd}

This case is more complicated than the even $m$ case. We have two possible fusion rules in (\ref{fusion2}); further more, the second fusion rule in (\ref{fusion2}) implies non-Abelian statistics. Though more complicated, one can analyze the braiding statistics using very similar arguments as in the even-$m$ case.

We begin with the first fusion rule in (\ref{fusion2}), $\xi\times\bar\xi=0$. The analysis for this case is almost identical to that for even $m$, so we only briefly describe the derivation. First, this fusion rule implies that all excitations are Abelian. Next, we define the topological invariant,
\begin{equation}
\Theta = m \theta_\xi \label{inv_sup1}
\end{equation}
Note that  (\ref{inv_sup1}) and (\ref{invariant}) differ by a factor of 2. Despite of this difference, $\Theta$ holds the same two properties as in the even-$m$ case: (i) it only depends on the flux of $\xi$; (ii) the full set of braiding statistics can be reconstructed out of $\Theta$.
They can be proved following similar arguments as in the even-$m$ case, and we do not repeat here. One can show that there exists a special vortex $\xi_0$ that carries unit flux and that has $\theta_{\xi_0} = \frac{\Theta}{m}$. A general excitation  can be obtained by fusing $k$ copies of $\xi_0$'s and further fusing a charge $q$. We label it as $(\xi_0; k, q)$. Its exchange statistics is given by
\begin{equation}
\theta_{(\xi_0;k,q)} =  \frac{\Theta}{m}k^2 + \frac{\pi}{m} kq + \pi q \label{ss1}
\end{equation}
where $k,q$ take values in the range $0, \dots, (2m-1)$. Finally, it can be shown that $\Theta$ takes values in the following form
\begin{equation}
\Theta = \frac{\pi}{4m} p \label{constraint21}
\end{equation}
where $p$ is an integer. We see that $\Theta$ can take $8m$ different values when $p$ runs in the range $0,1,\dots, 8m-1$.

The second fusion rule in (\ref{fusion2}) leads to non-Abelian statistics. Indeed, as implied by the fusion rule, $\xi$ is a non-Abelian anyon with quantum dimension $d_\xi = \sqrt 2$. Nevertheless, we can still define the topological invariant $\Theta$ as in (\ref{inv_sup1}). Note that since $\xi$ is non-Abelian, $\theta_\xi$ is now the topological spin of $\xi$. It can be shown that $\Theta$ still holds the same two properties as before. Since it is technical to show the two properties with non-Abelian statistics, we have moved the detailed derivation to  Appendix \ref{app2}. In Appendix \ref{app2}, we are able to find $3m^2$ excitations in total, a labeling scheme of the excitations, their fusion rules, and their topological spins from a given $\Theta$. 
We also find that possible values of $\Theta$ still take the form (\ref{constraint21}), except that $p$ can only be half integers $\frac{1}{2}, \frac{3}{2}, \dots, \frac{16m-1}{2}$. Again, there are $8m$ distinct values of $\Theta$ in this case.

Combining the braiding statistics from both fusion rules, we see that distinct types of braiding statistics are labeled by distinct values of $\Theta$, and $\Theta$ takes values in the following form
\begin{equation}
\Theta = \frac{\pi}{4m} p \label{constraint2}
\end{equation}
where $p$ runs in the range $0,\frac{1}{2}, 1, \frac{3}{2}, \dots, \frac{16m-1}{2}$. Integer values of $p$ correspond to Abelian statistics and the first fusion rule in (\ref{fusion2}),  and half-integer values of $p$ correspond to non-Abelian statistics and the second fusion rule in (\ref{fusion2}). We see that there are $16m$ distinct types of braiding statistics in total. Again, we show in Sec.~\ref{sec3} that all the $16m$ types of braiding statistics can be physically realized.

\subsection{Comparison to bosonic matter}

In passing, we briefly compare braiding statistics in gauge theories coupled fermionic matter versus bosonic matter. The above analysis for $\mathbb Z_{2m}$ gauge theories coupled fermionic matter can be easily adapted to $\mathbb Z_{2m}$ discrete gauge theories coupled to bosonic matter. For bosonic matter, all charge excitations are bosons. That is, Eq.~(\ref{exchange}) should be replaced by $\theta_q=0$. With this replacement, we then go through the same arguments as above, except two differences: (i) The constraint (\ref{constraint}) still holds, but the only solution is  $q=0$; (ii) The topological invariant $\Theta$ can be defined as in (\ref{invariant}) and takes values in the form (\ref{constraint1}), but $p$ has to be even integers.  The second point follows the fact that the exchange statistics of the collection of $2m$ $\xi$'s must be bosonic. Hence, there are $2m$ types of braiding statistics for both even and odd $m$, which is  fewer than the case of fermionic matter.



\section{Model realization}
\label{sec3}

To complete the analysis for braiding statistics in gauged charge-$2m$ superconductors, we now construct two toy models to realize all types of braiding statistics found in Sec.~\ref{sec2}.  Interestingly, the construction only involves free fermions with weak perturbations from interaction.

The first model is constructed to realize all the Abelian statistics, for both even and odd $m$. To construct the model,  we begin with an integer quantum Hall (IQH) system at a filling factor $\nu$, which has a charge $U(1)$ symmetry and a bulk energy gap. If we adiabatically  insert a $U(1)$ flux $\phi$ in the bulk,  the exchange statistics $\theta$ of the flux is given by
\begin{equation}
\theta= \frac{\phi^2}{4\pi}\nu \label{iqh}
\end{equation}
which can be found using the standard Chern-Simons theory\cite{wen-book}. Next, we imagine breaking the $U(1)$ symmetry to $\mathbb Z_{2m}^f$ by introducing a weak perturbation, e.g., by putting the system in proximity to another charge-$2m$ superconductor. In this way, we have turned the IQH system into a charge-$2m$ superconductor. We require the perturbation to be weak enough so that the energy gap does not close. Because of that, we expect the statistics $\theta$ does not change, except that now the flux $\phi$ is quantized to $\frac{\pi}{m}k$. Taking $\phi$ to be the unit flux $\frac{\pi}{m}$, we immediately obtain that the topological invariant $\Theta = 2m \theta = \frac{\pi}{2m}\nu$ for even $m$, and $\Theta = m\theta = \frac{\pi}{4m}\nu$ for odd $m$. By varying the filling factor $\nu$, we can exhaust all possible Abelian statistics.

The second model is only for odd $m$ and is constructed to realize the non-Abelian statistics. The construction relies on the fact that $\mathbb Z_{2m}^f = \mathbb Z_2^f \times \mathbb Z_m$ if $m$ is odd, where $\mathbb Z_2^f$ is the fermion parity. Let $a,b$ be the generators of $\mathbb Z_2^f$ and $\mathbb Z_m$ respectively, with $a^2=b^m=1$. That means, the generator of $\mathbb Z_{2m}^f$ is $ab$. To construct the model, we take two flavors of fermions, $\psi_1$ and $\psi_2$. Let them transform as follows:
\begin{align}
\psi_1 \overset{a}{\longrightarrow}-\psi_1, \quad &\psi_1 \overset{b}{\longrightarrow} \psi_1 \nonumber\\
\psi_2 \overset{a}{\longrightarrow}-\psi_2, \quad &\psi_2 \overset{b}{\longrightarrow} e^{-i\frac{2\pi}{m}}\psi_2
\end{align}
That is, the $\mathbb Z_m$ part acts trivially on $\psi_1$. Since $\psi_1$ only has a $\mathbb Z_2^f$ symmetry, we can put it into a $p_x+ip_y$ state. At the same time, we put $\psi_2$ in a state described by the first model with filling factor $\nu$.  The two fermions are completely decoupled.  We would like to know the topological spin of a $\mathbb Z_{2m}^f$ unit flux. Since $\psi_1$ sees the $\mathbb Z_{2m}^f$ unit flux as if it is a $\mathbb Z_{2}^f$ flux, it contributes $\frac{\pi}{8}$ to the topological spin---this is the property of $p_x+ip_y$ state\cite{kitaev06}. The contribution from $\psi_2$ is given by (\ref{iqh}). Combining the two contributions, we find that the topological invariant $\Theta$ is given by
\begin{equation}
\Theta = m\left(\frac{\pi}{8} + \frac{\pi}{4m^2}\nu\right) = \frac{\pi}{4m}\left(\nu+\frac{m^2}{2}\right)
\end{equation}
By varying the filling factor $\nu$, we are able to exhaust all possible non-Abelian statistics.


\section{Implications for classification of SPT phases}
\label{sec4}

We now use the results of braiding statistics to classify topological phases of {\it bare} charge-$2m$ superconductors (i.e., ungauged superconductors). Generally speaking, a charge-$2m$ superconductor may support chiral edge modes. Only those with nonchiral edge modes  belong to SPT phases.  A chiral superconductor is sometimes called {\it invertible topological phases}\cite{freed14}. The chirality of edge modes is characterized by the chiral central charge $c$, which is equivalent to the thermal Hall conductance\cite{kane97}.

Hence, we have two quantities, $\Theta$ and $c$, to characterize charge-$2m$ superconductors. Chiral central charge $c$ is invariant under any smooth deformation of the Hamiltonian of a charge-$2m$ superconductor as long as the energy gap does not close. Moreover, the topological invariant $\Theta$ defined for {\it gauged} superconductors is also invariant under smooth deformations of {\it bare} charge-$2m$ superconductors.  This can be guaranteed by choosing a proper gauging procedure (see Refs.~\onlinecite{levin12, wangcj15} for details): as long as the energy gap in the bare superconductor does not close, so does the energy gap of the gauged superconductor. As long as energy gap in the gauged superconductor does not close, $\Theta$ and the whole set of braiding statistics do not change. Hence, both $\Theta$ and $c$ are invariant under eligible smooth deformations of  charge-$2m$ superconductors. 

One way to classify topological phases is to find a {\it complete} set of physical quantities that are invariant under eligible smooth deformations. A complete set distinguishes every phase under consideration. For charge-$2m$ superconductors, we have the set $(\Theta,c)$.  While we cannot prove the completeness of $\Theta$ and  $c$, there is evidence from other studies\cite{levin12,gu14b,wangcj15} showing that they might form a complete set. Below we classify charge-$2m$ superconductors, based on the assumption that the data $(\Theta, c)$ is complete.

To proceed, we discuss properties of the data $(\Theta, c)$. The first property is that they are additive under stacking of two charge-$2m$ superconductors. More precisely, if we have two phases described by $(\Theta_1,c_1)$ and $(\Theta_2,c_2)$, stacking them together generates a new phase described by $(\Theta_1+\Theta_2, c_1+c_2 )$. Roughly speaking, $c$ counts the number of degrees of freedom on the edge, so it is not hard to see the additivity. To see the additivity of $\Theta$, we recall that $\Theta$ is a Berry phase associated with a vortex. Then, if we stack two systems, the total Berry phase associated with a vortex should be the sum of the Berry phases from each system.

The second property is that $\Theta$ and $c$ are not independent of one another. They are related through the following formula from general algebraic theory of anyons\cite{kitaev06}:
\begin{equation}
e^{i2\pi c/8 } = \frac{1}{D} \sum_\alpha d_\alpha^2e^{i\theta_\alpha}, \ D = \sqrt{\sum_\alpha d_\alpha^2}
\end{equation}
where $d_\alpha, \theta_\alpha$ are the quantum dimension and topological spin of $\alpha$, and the summations are over all anyons $\alpha$ in the theory. Applying this formula to our case, we find that
\begin{equation}
\begin{array}{lc}
\vspace{4pt}
\text{if $p$ is an integer:} & \ c \equiv p \modulo{8}\\
\text{if $p$ is a half integer:} & \ c\equiv p- \frac{m^2-1}{2} \modulo{8} \label{theta_c_relation}
\end{array}
\end{equation}
where $p$ parameterizes $\Theta$ through (\ref{constraint1}) for even $m$, and through (\ref{constraint2}) for odd $m$. To derive (\ref{theta_c_relation}), we have used (\ref{spin1}), (\ref{ss1}), and (\ref{ts1}) and (\ref{ts2}) from Appendix \ref{app2}. One can see that $c$ and $p$ are not mutually determined. There is a ``mod $8$'' uncertainty. This uncertainty is compensated by the following fact: there exists a state, called $E_8$ state, built out of fermion pairs, which has $c=8$ and $\Theta=0$\cite{e8} (See the discussion in Appendix \ref{app3}). Therefore, by stacking multiple copies of the $E_8$ state or its time reversal, we can shift $c$ by a multiple of $8$ while keeping $p$ unchanged.

With the above properties of $(\Theta,c)$, we are now ready to classify charge-$2m$ superconductors. It is clear that there should be two generating phases: (i) the phase described by $(\Theta^*, 0)$,  where $\Theta^*$ is the smallest nonzero value of $\Theta$ that is compatible with $c=0$; (ii) the phase described by $(\Theta_{c^*}, c^*)$, where $c^*$ is the smallest positive chiral central charge and $\Theta_{c^*}$ is any value of $\Theta$ that is compatible with $c^*$. According to (\ref{theta_c_relation},\ref{constraint1},\ref{constraint2}), we find that
\begin{align}
\text{$m$ being even:} &\quad \Theta^* = \frac{4\pi}{m}, \ c^* =1, \ \ \Theta_{c^*} = \frac{\pi}{2m}  \nonumber \\
\text{$m$ being odd:}\ &\quad \Theta^* =  \frac{2\pi}{m},  \ c^* = \frac{1}{2},\  \Theta_{c^*} = \frac{m\pi}{8}  \label{generators}
\end{align}
Note that the choice of $\Theta_{c^*}$ is not unique. Other phases can be obtained through stacking of the two generating phases.

The first generating phase generates phases described by $(t\Theta^*,0)$, where $t=0,1,\dots, t_{\rm max}$. Here, $t_{\rm max}=\frac{m}{2}-1$ for even $m$ and $t_{\rm max} = m-1$ for odd $m$. These phases all have $c=0$, which can be interpreted as fermionic SPT phases. The second generating phase (and its time reversal) generates phases described by $(r\Theta_{c^*},rc^*)$, where $r$ is any integer. These phases are associated with non-vanishing $c$.  Combining both types of phases, we obtain all possible invertible topological phases with $\mathbb Z_{2m}^f$ symmetry. The group structure of these phases under stacking is summarized in Table \ref{tab1}.

\begin{table}[t]
\caption{Classification of 2D fermionic SPT phases and invertible topological phases with $\mathbb Z_{2m}^f$ symmetry, based on the data $(\Theta, c)$. The ``$c^*$'' column shows the chiral central charge of the generating phase for each $\mathbb Z$ component. }\label{tab1}
\begin{tabular}{c|cccc}
\hline\hline
 &   $\quad $SPT $\quad$  & $\quad  $ Invertible $\quad $ & $\quad c^* \quad $ \\
\hline
$\quad $ even  $m$  $\quad $  & $\quad \mathbb Z_{m/2}\quad $ &  $\ \ \mathbb Z\times \mathbb Z_{m/2}\ \ $ &$\quad 1 \quad $  \\
$\quad $ odd  $m$ $\quad $   & $\quad \mathbb Z_m \quad $ & $\mathbb Z \times \mathbb Z_{m}$ & $\quad 1/2\quad $  \\
\hline
\end{tabular}
\end{table}


Finally, we discuss two interesting observations  from Table \ref{tab1} for the experimentally most relevant example, the charge-4 superconductors. First, we observe that there is no nontrivial $\mathbb Z_4^f$ fermionic SPT phase. This contradicts with the claim of Ref.~\onlinecite{lu12}, but agrees with the formal cobordism analysis.\cite{kapustin14}  In Appendix \ref{app4}, we further verify this point by an alternative analysis from edge theory. Second, all charge-4 superconductors have $c$ being integers. That means,  if we break $\mathbb Z_{4}^f$ symmetry down to $\mathbb Z_2^f$ in any phase, we can never obtain a $p_x+ip_y$ state, since the latter has $c=1/2$.
Consequently, $p_x+ip_y$ superconductors are intrinsically {\it incompatible} with $\mathbb Z_{4}^f$ symmetry. This property holds for any symmetry group that contains $\mathbb Z_4^f$ as a subgroup, including $\mathbb Z_{2m}^f$ with even $m$ and charge $U(1)$ symmetry.

\section{Conclusion}
\label{conclusion}

To summarize, we derive a classification of braiding statistics in charge-$2m$ superconductors that are coupled to a dynamical gauge field. We show that there exist $16m$ types of braiding statistics when $m$ is odd, while there are only $4m$ types when $m$ is even. Based on braiding statistics, we also obtain a classification of 2D fermionic SPT phases as well as invertible topological phases with $\mathbb Z_{2m}^f$ symmetry.  We envision a generalization of our method to 2D gapped fermionic systems with a general Abelian symmetry.

\begin{acknowledgments}
C.W. thanks  Z.-C. Gu, T. Lan, C.-H. Lin, M. Metlitski, X.-G. Wen, and M. Levin for enlightening discussions. In particular, C.W. thanks M. Levin for a careful reading of the manuscript and for encouragement and suggestions. This research was supported in part by Perimeter Institute for Theoretical Physics. Research at Perimeter Institute is supported by the Government of Canada through the Department of Innovation, Science and Economic Development Canada and by the Province of Ontario through the Ministry of Research, Innovation and Science.
\end{acknowledgments}

\appendix

\section{Proof of Eq.~(\ref{braid1})}
\label{app1}

\begin{figure}[b]
\centering
\includegraphics{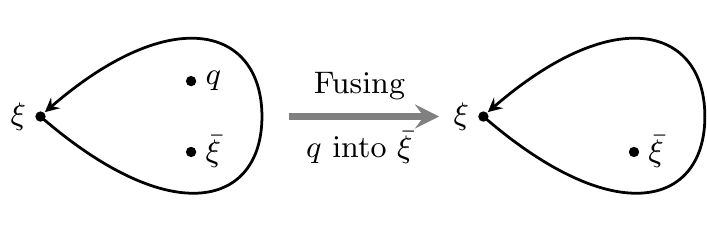}
\caption{Thought experiment to show Eq.~(\ref{braid1}) }\label{fig1}
\end{figure}

In this appendix, we prove Eq.~(\ref{braid1}) in the main text. To do that, we consider the thought experiment in Fig.~\ref{fig1}. In this thought experiment, we consider a state that contains three excitations, $\xi$, $\bar\xi$ and $q$, where $\xi$ is a vortex carrying unit flux $\frac{\pi}{m}$, $\bar\xi$ is its anti-particle and $q$ is a charge appearing in the fusion product $\xi\times\bar\xi$. In the initial state, $\xi$ and $\bar\xi$ are in the vacuum fusion channel. We imagine braiding $\xi$ around both $\bar\xi$ and $q$. This braiding process can be divided into two steps: we first braid $\xi$ around $\bar\xi$, then braid $\xi$ around $q$. The first step leads to a statistical phase $R_{\bar\xi\xi}^0R_{\xi\bar\xi}^0$, and the second step leads to a statistical phase $e^{i\pi q/m}$. The latter follows from the Aharonov-Bohm law (\ref{ab_phase}). Therefore, the overall statistical phase is given by $R_{\bar\xi\xi}^0R_{\xi\bar\xi}^0e^{i\pi q/m}$.

Yet, there is an alternative way to compute the overall statistical phase. Since we require $q$ to be one of the possible fusion channels of $\xi$ and $\bar\xi$, we can absorb $q$ into $\bar\xi$, at the same time leaving $\bar\xi$ unchanged (Fig.~\ref{fig1}). This follows from the fact that the fusion multiplicities satisfy $N_{\xi\bar\xi}^q = N_{\bar\xi\xi}^{\bar q}= N_{\bar\xi q}^{\bar \xi} = 1$, and thereby $q\times\bar\xi = \bar\xi$. One can match quantum dimensions on the two sides to see that only $\bar\xi$ appears on the right-hand side. What changes by the absorption is the fusion channel between $\xi$ and $\bar\xi$: they were in the vacuum channel, and now they are in the channel $q$. After the absorption, braiding $\xi$ around $\bar\xi$ gives the statistical phase $R_{\bar\xi\xi}^q R_{\xi\bar\xi}^q$. Importantly, we notice that the absorption process commutes with the braiding process, because paths of the two processes do not overlap. Therefore, the statistical phase associated with the braiding process before the absorption is also given by $R_{\bar\xi\xi}^q R_{\xi\bar\xi}^q$.

Combining the two ways of computation, Eq. (\ref{braid1}) immediately results. We comment that the roles of $\xi$ and $\bar\xi$ can be exchanged in the above thought experiment. A consequence is that we can replace $q$ with $-q$ on the right-hand side of (\ref{braid1}), and the equation still holds.

\section{Details of non-Abelian statistics for odd $m$}

\label{app2}

In this appendix, we discuss the details of braiding statistics for the second fusion rule, $\xi\times\bar\xi = 0 + Q$, in Eq.~(\ref{fusion2}). We show that the topological invariant $\Theta$, defined in (\ref{inv_sup1}), satisfies the following properties: (i) it is independent of the choice of $\xi$ but only depends on the flux of $\xi$; (ii) the full set of braiding statistics data can be reconstructed out of a given $\Theta$.  At the end, we also obtain all consistent values that $\Theta$ can take.

To show point (i), let us imagine replacing $\xi$  with $\xi'$ in the definition (\ref{inv_sup1}), where $\xi'$ also carries unit flux $\frac{\pi}{m}$. Since $\xi,\xi'$ carry the same gauge flux, they only differ by some charge. In other words, $\xi'$ can be represented as $\xi\times q$ for some charge $q$. Next, we calculate the difference in $\Theta$, resulting from the replacement. To do that, we use the following formula from the algebraic theory of anyons\cite{kitaev06}
\begin{align}
R_{\beta\alpha}^\gamma R_{\alpha\beta}^\gamma & = e^{i(\theta_\gamma-\theta_\alpha-\theta_\beta)} {\rm id}_{\mathbb V_{\alpha\beta}^\gamma} \label{topospin2}
\end{align}
where $R_{\alpha\beta}^\gamma$ is the $R$ symbol associated with a half braiding between $\alpha$ and $\beta$ in the fusion channel $\gamma$, ${\rm id}_{\mathbb V_{\alpha\beta}^\gamma}$ is the identity matrix in the fusion space $\mathbb V_{\alpha\beta}^\gamma$ (note that Eq.~(\ref{topospin}) is a special case of Eq.~(\ref{topospin2})). Making the substitutions $\alpha\rightarrow \xi$, $\beta\rightarrow q$ and $\gamma \rightarrow \xi'$, we find that
\begin{equation}
\theta_{\xi'} = \theta_{\xi}  + \frac{\pi q}{m} + \pi q \label{attachcharge}
\end{equation}
where we have used the fact that the mutual statistics $R_{q\xi}^{\xi'} R_{\xi q}^{\xi'}$ between $q$ and $\xi$ is given by the Aharonov-Bohm phase. We then immediately have $m\theta_{\xi'} = m\theta_{\xi}$, i.e., the topological invariant $\Theta$ only depends on the flux of $\xi$.

The rest of this section is devoted to showing point (ii), i.e., reconstructing the braiding statistics data out of a given $\Theta$. The braiding statistics data that we will derive includes a labeling scheme for the excitations, their fusion rules, and their topological spins. This set of braiding statistics data is equivalent to the commonly used $S$ and $T$ matrices.  It is conjectured that it is also equivalent to the $F$ and $R$ symbols for anyons described by unitary modular tensor category; recent progress on this conjecture can be found in Ref.~\onlinecite{lan15}. (In the case of Abelian anyons, this conjecture can be proved.) It is plausible that in our case, we can derive the $F$ and $R$ symbols for some gauge choice from the fusion rules and topological spins, but we do not try to obtain them since they are not gauge invariant (physical) quantities. General definition of topological spin $\theta_\alpha$ of an anyon $\alpha$ will be useful later, so we give it here\cite{kitaev06}:
\begin{equation}
e^{i\theta_\alpha} = \frac{1}{d_\alpha}\sum_{\gamma} d_\gamma {\rm Tr}(R_{\alpha\alpha}^\gamma) \label{spindef}
\end{equation}
where $d_\alpha,d_\gamma$ are quantum dimensions of $\alpha$ and $\gamma$, $R_{\alpha\alpha}^\gamma$ is the braid matrix associated with a half braiding of two $\alpha$'s, and the summation is over all fusion channels in $\alpha\times\alpha$.

To begin, we construct the braiding statistics of vortices that carry unit flux from a given $\Theta$. Let us again use $\xi$ to denote a general vortex that carries unit flux  $\frac{\pi}{m}$. It satisfies the fusion rule $\xi\times\bar\xi = 0 + Q$, where $Q$ is a fermion carrying gauge charge $m$. The first consequence that we can draw from this fusion rule is that the quantum dimension $d_\xi = \sqrt 2$. It follows from that $d_\xi = d_{\bar\xi}$, $d_0=d_Q=1$, and $d_{\xi} d_{\bar\xi} = d_0+d_Q$ which is a result of general anyon theory\cite{kitaev06}. Next, we determine the topological spin $\theta_\xi$ of any $\xi$ from a given $\Theta$. From the definition (\ref{inv_sup1}) of $\Theta$, we see that the topological spin $\theta_\xi$ should be in the form $\frac{\Theta}{m} + \frac{2\pi}{m}(\text{integer})$. If we attach a charge $q$ to $\xi$, we obtain another vortex whose topological spin is given by $\theta_\xi + \frac{\pi}{m}q + \pi q$, according to Eq.~(\ref{attachcharge}). Then, if we choose $q$ properly, we can obtain a special vortex $\xi_0$ such that
\begin{equation}
\theta_{\xi_0} = \frac{\Theta}{m} \label{xispin}
\end{equation}
With the vortex $\xi_0$, any other vortex carrying unit flux can be obtained by attaching some charge $q$ onto $\xi_0$. One can show that it is possible to obtain $m$ distinct vortices with $q=0,1,\dots, m-1$. The topological spins of these vortices are  $\frac{\Theta}{m} + \frac{\pi}{m}q + \pi q$, and are  different for different $q$. There is no other vortices that carry unit flux. Indeed, if we attach $Q$, i.e., $q=m$, to $\xi_0$, we obtain $\xi_0$ itself.

Next, we study vortices that carry twice of the unit flux. Let us use $\eta$ to denote a general vortex that carry flux $\frac{2\pi}{m}$. First, we consider the fusion rule of $\eta\times\bar\eta$. In general, $\eta\times\bar\eta = 0 + q+\dots$, with the right-hand side all being charges. Following a similar argument to the case of $\xi\times\bar\xi$, we find that the condition on which charge $q$ can appear in the fusion product $\eta\times\bar\eta$ is given by
\begin{equation}
e^{i\frac{2\pi}{m}q} = e^{i\pi q}
\end{equation}
For $m$ being odd, the only solution to this condition is $q=0$. Therefore, every $\eta$ is an Abelian anyon. Secondly, we consider the fusion product $\xi\times\xi$, where the fusion outcome should be some $\eta$'s. By matching quantum dimensions,  it is not hard to see that the only possible fusion rule is
\begin{equation}
\xi\times\xi=\eta_1+\eta_2
\end{equation}
where both $\eta_1,\eta_2$ carry gauge flux $\frac{2\pi}{m}$. In fact, $\eta_1, \eta_2$ are related through $\eta_1\times Q = \eta_2$. To see that, imagine we have two $\xi$'s in the fusion channel $\eta_1$. We then fuse $Q$ into one of the two $\xi$'s. Since $Q\times\xi=\xi$, this fusion process give two $\xi$'s in the fusion channel $Q\times\eta_1$. Since the topological spin of $Q\times \eta_1$ is given by $\theta_{\eta_1} + \pi$, it is a different anyon from $\eta_1$ and it must be $\eta_2$. This proves $\eta_2=Q\times \eta_1$.

What are the topological spins $\theta_{\eta_1}, \theta_{\eta_2}$ (or equivalently the exchange statistics, since $\eta$'s are Abelian)? We would like to relate $\theta_{\eta_1}, \theta_{\eta_2}$ to the topological spin $\theta_\xi$. To do that, we first use the definition (\ref{spindef}) of topological spin
\begin{equation}
e^{i\theta_\xi} = \frac{1}{\sqrt 2} (R_{\xi\xi}^{\eta_1} + R_{\xi\xi}^{\eta_2}) \label{na1}
\end{equation}
In our case, the fusion multiplicities are all equal to 1, so the $R$ symbols are complex numbers. At the same time, according to Eq.~(\ref{topospin2}), we have $(R_{\xi\xi}^{\eta_i} )^2 = e^{i\theta_{\eta_i} - i2\theta_{\xi}}$, where $i=1,2 $. Combining  with the fact that $\theta_{\eta_2} = \theta_{\eta_1} + \pi$,  we obtain $(R_{\xi\xi}^{\eta_1} )^2 +(R_{\xi\xi}^{\eta_2} )^2 = 0$. Further combining with (\ref{na1}), we obtain
\begin{equation}
 R_{\xi\xi}^{\eta_1}R_{\xi\xi}^{\eta_2}= \left(e^{i\theta_\xi}\right)^2 \label{na2}
\end{equation}
Equations (\ref{na1}) and (\ref{na2}) imply that $R_{\xi\xi}^{\eta_1} $, $R_{\xi\xi}^{\eta_2} $ are the two solutions to the quadratic equation
\begin{equation}
x^2 - \sqrt 2 w x +w ^2=0. \label{na3}
\end{equation}
where $x$ is the unknown and $w =e^{i\theta_\xi}$. Solving (\ref{na3}), we find
\begin{equation}
R_{\xi\xi}^{\eta_1} = e^{i\pi/4 + i\theta_{\xi}}, \ R_{\xi\xi}^{\eta_2} = e^{-i\pi/4 + i \theta_\xi }
\end{equation}
where we have chosen one of the solutions to be $R_{\xi\xi}^{\eta_1}$ without losing any generality. Therefore, we obtain the following topological spins
\begin{equation}
\theta_{\eta_1} =  \frac{\pi}{2} + 4\theta_\xi, \quad \theta_{\eta_2} = -\frac{\pi}{2} + 4 \theta_\xi
\end{equation}

With these results, let us choose $\eta_0$ to be one of the two fusion outcomes of $\xi_0\times\xi_0$, with the topological spin
\begin{equation}
\theta_{\eta_0} = \frac{\pi}{2} + \frac{4\Theta}{m} \label{etaspin}
\end{equation}
From $\eta_0$, one can show that any $\eta$ can be obtained by attaching some charge $q$ onto $\eta_0$, and the topological spin of the resulting vortex is equal to $\theta_{\eta_0} + \frac{2\pi}{m}q +  \pi q$. Varying $q$ in the range $0,\dots, (2m-1)$, we are able to obtain $2m$ distinct vortices that carry flux $\frac{2\pi}{m}$.

Before moving on to vortices that carry other gauge flux, we would like to study the mutual statistics $\theta_{\eta_0\xi_0}$ between the two vortices  $\eta_0$ and $\xi_0$. It is an Abelian phase because $\eta_0$ is Abelian. To do that, we instead first compute the mutual statistics $\theta_{\eta_0\bar\xi_0}$ between $\eta_0$ and $\bar\xi_0$. Since $  N_{\eta_0\bar\xi_0 }^{\xi_0} = N_{\bar\xi_0\bar\xi_0}^{\bar\eta_0}= N_{\xi_0\xi_0}^{\eta_0} = 1$, we have the fusion rule $\eta_0 \times \bar\xi_0 = \xi_0$. Therefore, using the relation (\ref{topospin2}), the mutual statistics $\theta_{\eta_0\bar\xi_0}$ can be expressed as
\begin{equation}
e^{i\theta_{\eta_0\bar\xi_0}}= e^{i\theta_{\xi_0} - i\theta_{\bar\xi_0} -i \theta_{\eta_0}} = e^{-i\theta_{\eta_0}}
\end{equation}
where we have used the property that $\theta_{\xi_0} = \theta_{\bar\xi_0}$ \cite{kitaev06}. To proceed, we consider braiding $\eta_0$ around both $\xi_0$ and $\bar\xi_0$. One can see that regardless of whether $\xi_0$ and $\bar\xi_0$ are in the $0$ or $Q$ fusion channel, the statistical phase is always $0$. Therefore, $\theta_{\eta_0\xi_0} + \theta_{\eta_0\bar\xi_0}=0$. Hence, the mutual statistics $\theta_{\eta_0\xi_0}$ is given by
\begin{equation}
\theta_{\eta_0\xi_0} = \theta_{\eta_0}\label{mutual}
\end{equation}

We are now ready to work out the general case. A general vortex carries gauge flux $\frac{\pi}{m} k$, where $k=1,\dots, (2m-1)$.  First, let us construct the vortices with even $k$. This can be done by fusing $\frac{k}{2}$ copies of $\eta_0$'s and a charge $q$. That is,
\begin{equation}
(\xi_0, \eta_0; k, q) = \overbrace{\eta_0 \times \dots \times \eta_0}^{k/2 \text{ times}} \times q \label{def1}
\end{equation}
where we have used the notation $(\xi_0, \eta_0; k,q)$ to denote the vortex. (We think of $\xi_0, \eta_0$ as {\it reference vortices}, so that they are included in our notation.) Obviously, this vortex carries gauge flux $\frac{\pi}{m}k$, and it is an Abelian anyon. Its topological spin can be computed by  using (\ref{topospin2}) recursively and by using the topological spin (\ref{etaspin}) of $\eta_0$,  and is given by
\begin{equation}
\theta_{(\xi_0, \eta_0; k,q)} = \left(\frac{\pi}{8} + \frac{\Theta}{m}\right)k^2 + \frac{\pi }{m} kq + \pi q \label{ts1}
\end{equation}
where $k$ is even. For a fixed $k$, the value of $q$ can be $0, 1,\dots, (2m-1)$. All these vortices are distinct, which can be verified by checking the fact that  they either have different topological spins or different mutual statistics with respect to $\eta_0$. There are no other vortices carrying gauge flux $\frac{\pi}{m}k$ with even $k$.

Next, we construct the vortices with odd $k$. This can be done by fusing a $\xi_0$,  $\frac{k-1}{2}$ copies of $\eta_0$'s,  and a charge $q$. That is,
\begin{equation}
(\xi_0, \eta_0; k, q) =\xi_0\times \overbrace{\eta_0 \times \dots \times \eta_0}^{(k-1)/2 \text{ times}} \times q \label{def2}
\end{equation}
We note that the fusion on the right-hand side indeed gives a {\it single} anyon. The vortex $(\xi_0, \eta_0; k, q)$ carries gauge flux $\frac{\pi}{m}k$, and it is non-Abelian. The topological spin can be computed with (\ref{topospin2}), (\ref{xispin}), (\ref{etaspin}) and (\ref{mutual}), and is given by
\begin{equation}
\theta_{(\xi_0,\eta_0; k,q)} = \left(\frac{\pi}{8} + \frac{\Theta}{m}\right)k^2 + \frac{\pi }{m} kq + \pi q -\frac{\pi}{8} \label{ts2}
\end{equation}
where $k$ is odd. For a fixed $k$, the value of $q$ can be $0, 1, \dots, (m-1)$. All these vortices are distinct, which can be verified by checking the fact that they have distinct mutual statistics with respect to $\eta_0$. There are no other vortices that carries gauge flux $\frac{\pi}{m}k$ with odd $k$.

This completes our search of vortices. We see that the total number of excitations, including charges, is $3m^2$. Their topological spins are given by (\ref{ts1},\ref{ts2}).

Finally, we still need to find the fusion rules to complete the construction of our braiding statistics data. General fusion rules can be found by using the definitions (\ref{def1}), (\ref{def2}) and some basic fusion rules involving $\xi_0, \eta_0$ and charges. We collect the basic fusion rules below:
\begin{align}
\xi_0\times\xi_0 &= \eta_0 +\eta_0\times Q \label{f1}\\
\xi_0\times Q &= \xi_0 \label{f2}\\
q\times q' & = [q+q'] \label{f3}\\
\overbrace{\eta_0\times\dots\times\eta_0}^{\text{$m$ times}} & = q^*  \label{f4}
\end{align}
Several comments about these fusion rules are as follows. In (\ref{f1}), $\eta_0\times Q$ represents the vortex obtained by fusing $\eta_0$ and $Q$. In (\ref{f3}), $q,q'$ are general charge excitations and $[q+q']$ means the residue of $q+q'$ modulo $2m$. The fusion rule (\ref{f4}) is worth to paying special attentions. The excitation $q^*$ is a charge determined by $\Theta$:
\begin{equation}
q^* = \frac{m^2}{2} + \frac{4 m\Theta}{\pi} \label{constraint_na}
\end{equation}
To see that, we imagine braiding $\xi_0$ around $m$ copies of $\eta_0$'s. Using Eq.~(\ref{mutual}), we see the statistical phase equals $m\theta_{\eta_0} = m\pi/2 + 4\Theta$. On the other hand, we know $m$ $\eta_0$'s must be fused to a charge, and we denote the charge as $q^*$. Then, the statistical phase is also equal to $\frac{\pi}{m} q^*$. The two ways of computation must give the same result, thereby leading to (\ref{constraint_na}).  Interestingly, since $q^*$ must be an integer, this relation constrains the value of $\Theta$. One can see that
\begin{equation}
\Theta =\frac{\pi}{4m}p \label{cc}
\end{equation}
where $p$ is a half integer. Distinct values of $\Theta$ can be obtained by taking  $p = \frac{1}{2}, \frac{3}{2},\dots, \frac{16m-1}{2}$. There are $8m$ distinct values in total.

With the above basic fusion rules, it is not hard to find general fusion rules. For example, consider the fusion rule of two vortices $(\xi_0,\eta_0; k_1, q_1)$ and $(\xi_0, \eta_0; k_2, q_2)$, with both $k_1,k_2$ being odd. One can find that if $k_1 + k_2 < 2m$,
\begin{align}
  &  (\xi_0,  \eta_0; k_1,  q_1)\times (\xi_0, \eta_0; k_2,  q_2) \nonumber \\
  & = \xi_0\times \overbrace{\eta_0 \times \dots \times \eta_0}^{(k_1-1)/2 \text{ times}} \times q_1\times \xi_0\times \overbrace{\eta_0 \times \dots \times \eta_0}^{(k_2-1)/2 \text{ times}} \times q_2 \nonumber\\
   & =  \overbrace{\eta_0 \times \dots \times \eta_0}^{(k_1+k_2)/2 \text{ times}} \times q_1\times q_2 + \overbrace{\eta_0 \times \dots \times \eta_0}^{(k_1+k_2)/2 \text{ times}} \times q_1\times q_2\times Q \nonumber\\
 &  = (\xi_0, \eta_0; k_1+ k_2, [q_1+q_2]) \nonumber\\
 & \quad + (\xi_0, \eta_0; k_1+ k_2, [q_1+q_2+Q])
\end{align}
where the commutativity of fusion has been used. If $k_1+k_2 \ge 2m$, it can be similarly shown that
\begin{align}
  & (\xi_0, \eta_0; k_1, q_1)\times (\xi_0, \eta_0; k_2, q_2)  \nonumber \\
 & =   (\xi_0, \eta_0; k_1+ k_2 -2m, [q_1+q_2 + q^*]) \nonumber \\
  &  \quad + (\xi_0,\eta_0; k_1+ k_2-2m, [q_1+q_2+q^* + Q])
\end{align}
Other fusion rules can be similarly obtained. This completes the reconstruction of the braiding statistics data from a given $\Theta$.


\section{$E_8$ state in fermion systems}
\label{app3}

The purpose of this appendix is to show that there exists an $E_8$ state built out of fermion pairs, which has the following properties: (1) it has $c=8$; (2) it has a $\mathbb Z_{2m}^f$ symmetry; (3)  the topological invariant $\Theta$ associated with this state is 0.

We construct such a state within the $K$-matrix formalism\cite{wen-book}. We start with a state, whose edge Lagrangian has the following $K$-matrix description\cite{wen-book, e8, lu12}
\begin{equation}
L = \frac{1}{4\pi} K_{IJ} \partial_t\Phi_I\partial_x\Phi_J -  V_{IJ}\partial_x\Phi_I\partial_x\Phi_J \label{edge1}
\end{equation}
where
\begin{equation}
K = \left(\begin{matrix}
2 & 1 & 0 & 0 & 0 & 0 & 0 & 0 \\
1 & 2 & 1 & 0 & 0 & 0 & 0 & 0 \\
0 & 1 & 2 & 1 & 0 & 0 & 0 & 0 \\
0 & 0 & 1 & 2 & 1 & 0 & 0 & 0 \\
0 & 0 & 0 & 1 & 2 & 1 & 0 & 1 \\
0 & 0 & 0 & 0 & 1 & 2 & 1 & 0 \\
0 & 0 & 0 & 0 & 0 & 1 & 2 & 0 \\
0 & 0 & 0 & 0 & 1 & 0 & 0 & 2
\end{matrix}\right)
\end{equation}
and $V$ is a non-universal velocity matrix, and $\Phi$ is an 8-component real field. The above $K$ matrix is exactly the one that describes the bosonic $E_8$ state\cite{lu12}. An additional feature that we add here is a charge $U(1)$ symmetry, under which the fields transform as
\begin{equation}
\Phi_I \rightarrow \Phi_I + \theta K^{-1}_{IJ} t_J
\end{equation}
where $\theta$ parameterizes the $U(1)$ angle, and the so-called charge vector $t$ equals
\begin{equation}
t^T=(0,0,0,0,0,0,0,0)
\end{equation}
The fact that $t=0$ means $U(1)$ acts trivially on $\Phi$.  Excitations in the state are described by $e^{il^T\Phi}$, where $l$ is any integer vector. The exchange statistics of an excitation is given by $\pi l^TK^{-1}l$. Since $\det(K)=1$ and all diagonal elements of $K$ are even, all excitations are bosons. Furthermore, all excitations are charge neutral. Therefore, this state is built out of neutral fermion pairs.

We now check that the state described by the above theory satisfies the three desired properties. First, this $K$ matrix has eight positive eigenvalues, so the chiral central charge $c=8$. Second, there is a $U(1)$ symmetry in the system. Since $U(1)$ contains $\mathbb Z_{2m}^f$ symmetry as a subgroup, $\mathbb Z_{2m}^f$ is also a symmetry of this state. Third, the filling factor of the above integer quantum Hall state is $\nu = t^TK^{-1}t =0$. Therefore, according to (\ref{iqh}), the topological invariant $\Theta$ also vanishes. Hence, we prove the existence of the desired $E_8$ state.

A remark is as follows. One may notice that there are no fermions in the theory. Nevertheless, we can always include fermions by adding some ``fermionic block'' to the $K$ matrix. For example, we can add $K_f = {\rm diag}(1,-1)$ to the matrix $K$, at the same time we add a piece $t_f^T=(1,1)$ to the charge vector $t$. We can show that adding these pieces introduces fermionic excitations, but does not change the above three properties.

\section{Trivialization of $\mathbb Z_2$ bosonic SPT state when embedded in $\mathbb Z_4^f$ fermion systems}
\label{app4}

In the main text, we argued through the braiding statistics approach that there is no nontrivial fermionic SPT phase with $\mathbb Z_4^f$ symmetry. However, one may intuitively construct a seemingly nontrivial state as follows: first let the fermions form strongly bound Cooper pairs, and then let the Cooper pairs form a nontrivial $\mathbb Z_2$ bosonic SPT state. (There are two $\mathbb Z_2$ bosonic SPT phases: one is trivial and the other is nontrivial.) This state was believed to be nontrivial by the authors of Ref.~\onlinecite{lu12}. However, we argue that such state is trivialized in the presence of fermions. Our argument is based on an analysis of edge stability/instability using the $K$-matrix formalism.\cite{wen-book}

To begin, let us be more precise on the symmetry transformations  of the fermions and the bosons (Cooper pairs). Let $g$ be the generator of $\mathbb Z_4^f$ symmetry with $g^4=1$. That is, $g^2$ is the fermion parity. Under the action of $g$, the fermion creation operator $\psi^\dag$ transforms as
\begin{equation}
\psi^\dag \overset{g}{\longrightarrow} e^{i\frac{\pi}{2}}\psi^\dag \label{edge00}
\end{equation}
Since bosons are thought of as Cooper pairs, the boson creation operator $b^\dag\sim \psi^\dag\psi^\dag$.  Then, it transforms as
\begin{equation}
b^\dag \overset{g}{\longrightarrow} e^{i\pi}b^\dag \label{edge0}
\end{equation}
Hence, the $\mathbb Z_4^f$ symmetry  acts as if it is a $\mathbb Z_2$ symmetry for bosons.

Bosons can form a nontrivial $\mathbb Z_2$ SPT state. In this state, there exist robust gapless edge modes that cannot be gapped out without breaking the $\mathbb Z_2$ symmetry. The edge modes can be described by the so-called $K$-matrix theory. A general $K$-matrix theory has a Lagrangian
\begin{equation}
L = \frac{1}{4\pi} K_{IJ} \partial_t\Phi_I\partial_x\Phi_J -  V_{IJ}\partial_x\Phi_I\partial_x\Phi_J \label{edge1}
\end{equation}
where $ K$  is a non-degenerate symmetric integer matrix, $V$ is the velocity matrix, and $\Phi$ is a multi-component field living in $1+1$ dimensions. For the nontrivial $\mathbb Z_2$ bosonic SPT phase, the edge theory can be described by a two-component field $\Phi$, with
\begin{equation}
K =\left(
\begin{matrix}
0 & 1\\
1 & 0
\end{matrix}
\right), \quad \Phi =
\left(
\begin{matrix}
\phi_1\\
\phi_2
\end{matrix}
\right)\label{edge2}
\end{equation}
while the velocity matrix $ V$ is irrelevant for the stability/instability analysis. Under the $\mathbb Z_2$ transformation, $\phi_1,\phi_2$ transform as follows
\begin{equation}
\phi_1 \overset{g}{\longrightarrow}\phi_1 + \pi, \quad \phi_2\overset{g}{\longrightarrow} \phi_2 +\pi \label{edge3}
\end{equation}
In this edge theory, the boson creation operators are of the form $b^\dag \sim e^{i\phi_1}$ or $b^\dag \sim e^{i\phi_2}$. Therefore, we see that the transformations (\ref{edge3}) are consistent with (\ref{edge0}). It was argued before that the edge theory described by (\ref{edge1}), (\ref{edge2}) and (\ref{edge3}) cannot be gapped out without breaking the $\mathbb Z_2$ symmetry\cite{levin12, lu12}.

We now analyze the stability of the above edge in the context of $\mathbb Z_4^f$ symmetric fermion systems. Since it is a fermion system now, we are allowed to enlarge  $K$ by adding some ``fermionic blocks''. We consider the following enlarged  theory,
\begin{equation}
\tilde{ K}=\left(
\begin{matrix}
0 & 1 & 0 & 0\\
1 & 0 & 0 & 0\\
0 & 0 & 1 & 0\\
0 & 0 & 0 & -1
\end{matrix}
\right), \quad \tilde\Phi =
\left(
\begin{matrix}
\phi_1 \\
\phi_2\\
\phi_3\\
\phi_4
\end{matrix}
\right) \label{edge4}
\end{equation}
where under the $\mathbb Z_4^f$ symmetry, the fields $\phi_3,\phi_4$ transform as
\begin{equation}
\phi_3 \overset{g}{\longrightarrow} \phi_3 + \frac{\pi}{2}, \quad \phi_4\overset{g}{\longrightarrow}\phi_4 - \frac{\pi}{2} \label{edge5}
\end{equation}
The fermion creation operators are $e^{i\phi_3}$ and $e^{-i\phi_4}$, which transform in the same way as in (\ref{edge00}). One property is that the fermionic block that we have added  is a trivial block, meaning that we can gap it out without breaking $\mathbb Z_4^f$ symmetry. One can check that if we add a perturbation $U\cos(\phi_3+\phi_4)$ to the edge, the edge modes $\phi_3,\phi_4$ will be gapped out at large $U$ without breaking the $\mathbb Z_4^f$ symmetry, and the edge theory will go back to the one described by (\ref{edge2}).

Our aim is to show that the edge described by $\tilde{ K}$ and $\tilde\Phi$, as well as the transformations (\ref{edge3}) and (\ref{edge5}), can be completely gapped out by suitable perturbations without breaking the $\mathbb Z_4^f$ symmetry. We find the following perturbations can do the job:
\begin{equation}
H'= U \cos\left(\Lambda_1^T\tilde{  K}\tilde \Phi \right) + U\cos\left(\Lambda_2^T\tilde{ K}\tilde \Phi\right)
\end{equation}
with
\begin{equation}
\Lambda_1 =
\left(
\begin{matrix}
1 \\
2\\
0\\
2
\end{matrix}
\right), \quad
\Lambda_2 =
\left(
\begin{matrix}
1 \\
0\\
1\\
1
\end{matrix}
\right) \label{edge6}
\end{equation}
First of all, we see that $H'$ is invariant under the transformations (\ref{edge3}) and (\ref{edge5}). Next, we need to show the perturbation $H'$ can gap out the edge at least for large $U$. It can be shown by using the so-called {\it null vector criterion}\cite{haldane95,levin12b}. According to that criterion, the edge will gap out for large $U$ if and only if
\begin{equation}
\Lambda_1^T \tilde { K }\Lambda_1 = \Lambda_1^T \tilde{  K }\Lambda_2= \Lambda_2^T \tilde{ K} \Lambda_2 =0
\end{equation}
One can easily check that the given $\Lambda_1$ and $\Lambda_2$ in (\ref{edge6}) indeed satisfy the null vector criterion. Finally, we still need to check that the edge does not break symmetry spontaneously after being gapped out. This can be done by checking the ground state degeneracy of the gapped edge: if there is a unique ground state, no spontaneous symmetry breaking occurs. For perturbations like $H'$, a general procedure to compute the  ground state degeneracy has been worked out in Refs.~\onlinecite{levin12b, wang13}. According to these works, the degeneracy is given by the greatest common divisor of all the $2\times 2$ minors of the matrix
\begin{equation}
(\Lambda_1, \Lambda_2) = \left(
\begin{matrix}
1 & 1 \\
2 & 0\\
0 & 1\\
2 & 1
\end{matrix}
\right)
\end{equation}
One can check that the ground state degeneracy is 1. So, there is no spontaneous symmetry breaking.

Hence, we have found a way to gap out the nontrivial $\mathbb Z_2$ bosonic SPT state by adding fermions. This completes our proof that the nontrivial $\mathbb Z_2$ bosonic SPT state is trivialized when it is embedded into a $\mathbb Z_4^f$ symmetric fermion system. This verifies the fact that there is no nontrivial SPT phase for $\mathbb Z_4^f$ symmetric fermions. One final remark is that if the nontrivial $\mathbb Z_2$ bosonic SPT state is embedded into a $\mathbb Z_2^f\times\mathbb Z_2$ fermion system, it is stable\cite{gu14b}.

\bibliography{spt}

\end{document}